\pdfoutput=1

\documentclass[twocolumn,showpacs, preprintnumbers,amsmath,amssymb]{revtex4-1}
\usepackage{dcolumn}% Align table columns on decimal point
\usepackage{bm}% bold math
\usepackage[dvips]{color} %removed for arxiv
\usepackage{ulem}
\usepackage[pdftex]{graphicx}
\usepackage{epstopdf}  % removed for arxiv
\usepackage{amsmath}

\providecommand{\tabularnewline}{\\}

\begin{document}

\preprint{APS/unknown}

\title{Probing La$_{0.7}$Sr$_{0.3}$MnO$_{3}$ multilayers via spin wave
resonances}

\author{Rhet Magaraggia}

\email{rhet.magaraggia@gmail.com}
\author{Mikhail Kostylev}
\author{Robert L Stamps}

\affiliation{School of Physics, University of Western Australia, 35 Stirling Highway,
Crawley, Western Australia 6009, Australia}

\author{Michael Hambe}
\author{Valanoor Nagarajan}
\affiliation{School of Materials Science and Engineering University of New South
Wales NSW 2052, Australia}

\begin{abstract}
La$_{0.7}$Sr$_{0.3}$MnO$_{3}$/BiFeO$_{3}$ and La$_{0.7}$Sr$_{0.3}$MnO$_{3}$/PbZr$_{20}$Ti$_{80}$O$_{3}$
epitaxial heterostructures have been grown on SrTiO$_{3}$ substrates. Spin wave resonances
are used to study interface properties of the ferromagnetic La$_{0.7}$Sr$_{0.3}$MnO$_{3}$.
We find that the addition of the BiFeO$_{3}$ or PbZr$_{20}$Ti$_{80}$O$_{3}$
causes out-of-plane surface pinning of the La$_{0.7}$Sr$_{0.3}$MnO$_{3}$.
We are able to place new limits on the exchange constant D of La$_{0.7}$Sr$_{0.3}$MnO$_{3}$
grown on these substrates and confirm the presence of uniaxial and
biaxial anisotropies caused by the SrTiO$_{3}$ substrate.
\end{abstract}
\maketitle

\section{Introduction}

The promise of electrically and magnetically tunable tunnel junctions
for use in both spin valves and four state memory devices\citep{springerlink:10.1007/s11434-008-0275-8,majumdar:122114,PhysRevB.75.245324},
is an exciting prospect. La$_{1-x}$Sr$_{x}$MnO$_{3}$/BiFeO$_{3}$(LSMO/BFO)
multilayers are proposed for spin valve devices, as LSMO has demonstrated
good spin filtering properties\citep{nature_LSMO} and BFO is a room
temperature multiferroic which could, in principle, provide an electrically
tunable exchange biased film\citep{doi:10.1021/nl801391m,adma.200502622,Nature-Mat-BFO-Bias}.
Both of these materials are perovskite based structures and have small
lattice mismatch when grown on a suitable substrate. Practically,
it is important to understand how the magnetization of LSMO films
is affected when epitaxially joined to a ferroelectric. Enhancement
of uniaxial anisotropies, development of unidirectional anisotropies,
surface pinning from the interface and changes to other micromagnetic
parameters are all important characteristics with respect to tunnel
junction performance. Though most of these properties have been explored
in single layer LSMO\citep{4837696,Gomes20101174,PhysRevB.78.094413,PhysRevB.76.184413,PhysRevB.73.054406,suzuki:140,lofland:1947,Dho200723,5026792}, the
effects of ferroelectric overlayers can be important and have begun to be studied\cite{PhysRevLett.104.167203,thiele:262502}. In this paper we examine pinning of
dynamic magnetization using spin wave measurements. A new result is
our measure for the spin wave exchange constant D. To the best of
the authors knowledge, only four other measurements of D have been
carried out so far\citep{PhysRevB.76.184413,PhysRevB.53.14285,vasiliu-doloc:7342,Moudden1997276},
and only one study which utilises standing spin wave modes for determination
of D\citep{PhysRevB.76.184413}. We also use standing spin wave resonances
to measure anisotropies caused by both the ferroelectric overlayer
and growth of LSMO on a single crystal substrate.

\section{Standing Spin Wave Modes}

A powerful technique to probe magnetic conditions at buried interfaces
is through spin wave resonances\citep{PhysRevB.38.6847,PhysRevB.58.8605,Kittel_Formula,Classic_Kittel_FMR,Spin_Wave_Resonance_First_Experiment}.

The structure of standing spin wave modes contains detailed information
about bulk magnetic properties, such as gyromagnetic ratio $\gamma$
and exchange constant D, and also provides information about interfacial
pinning of the magnetization vector.

The ferromagnetic resonance frequency for out-of-plane magnetized
thin films is

\begin{eqnarray}
\frac{\omega}{\gamma} & = & H_{eff}+H_{R}\label{eq:oopkittel-1}\end{eqnarray}

where $H_{eff}=-\mu_{0}M_{S}+H_{oop}+D\, k_{oop}^{2}$, $\omega$
is the precession frequency, $\gamma$ is the gyromagnetic ratio,
$H_{R}$ is the externally applied field , $\mu_{0}M_{S}$ is the
demagnetizing field due to the out-of-plane alignment of spins, $H_{oop}$
is any bulk out-of-plane anisotropy field and $D\, k_{oop}^{2}$ is
the exchange energy of the standing spin wave mode. Measurement of
multiple modes allows determination of $\gamma$ and $H_{eff}$. Separation of $-\mu_{0}M_{S}+H_{A}$ and
$D\, k_{oop}^{2}$, is possible when the fundamental resonance mode
frequency (FMR mode) and the first exchange mode (FEX mode) are measured.
Due to the shorter wavelength and much higher energy density of the
FEX mode, any changes in wavelength due to surface pinning strongly
affect the frequency gap between the FEX and FMR modes.

Subtraction of the effective field data of the FEX mode $H_{eff}(FEX)$,
from the FMR mode $H_{eff}(FMR)$ gives us a measure of the strength
of pinning and the exchange constant. By rotating the film in-plane
and taking angular measurements, the angular variation of pinning
and bulk anisotropies may be determined.

Angular dependence of spin wave frequencies for in plane magnetization
can be used to measure magnetocrystalline anisotropies. The Kittel
formula describing resonance conditions for the magnetization oriented
in-plane is\citep{Kittel_Formula}

\begin{eqnarray}
\left(\frac{\omega}{\gamma}\right)^{2} & = & (H_{R}(\theta)+\mu_{0}M_{S}-H_{oop}+H_{ip}(\theta)+D\, k_{ip}^{2}(\theta))\nonumber \\
 &  & \times(H_{R}(\theta)+H_{ip}(\theta)+D\, k_{ip}^{2}(\theta))\label{eq:kittel-1}\end{eqnarray}

Here $H_{ip}$ is the in-plane bulk anisotropy and $k_{ip}^{2}$refers
to the wave vector of standing wave modes with magnetization aligned
in plane. We make distinct $k_{ip}^{2}$ and $k_{oop}^{2}$
which need not in general be the same depending upon pinning conditions
at the film interface. Also included is an angular dependence $\theta$
which denotes the magnetization direction with respect to some arbitrary
in-plane film direction.

The paper is structured as follows. We first describe sample growth
and characterization, and the ferromagnetic resonance experiment.
FMR results are presented along with a discussion of bulk and surface
anisotropies. We conclude with results for the exchange constant D
of La$_{1-x}$Sr$_{x}$MnO$_{3}$ films.

\section{Sample growth and characterization}

\begin{figure}[th]
\begin{centering}
\includegraphics[width=9cm]{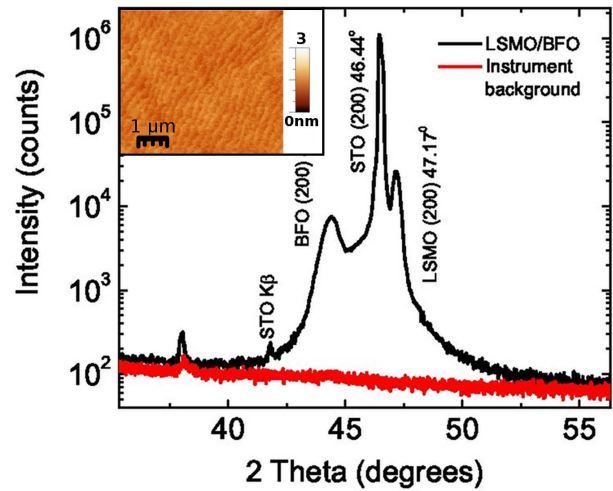}
\par\end{centering}

\caption{\label{fig:strucchar}A characteristic XRD image of the LSMO(55nm)/BFO(18nm)
film showing excellent phase purity. The inset demonstrates the step
structure imaged with AFM which originates on the surface of the of
the LSMO(38.9nm) sample due to epitaxial growth in on top of the stepped
substrate.}

\end{figure}

A series of films comprising epitaxial La$_{0.7}$Sr$_{0.3}$MnO$_{3}$ (LSMO) were grown on single crystal (100) orientated SrTiO$_{3}$ (STO) substrates, with the addition of either an epitaxial BiFeO$_{3}$ (BFO) or PbZr$_{20}$Ti$_{80}$O$_{3}$ (PZT) capping layer. As a comparision, a thick polycrystalline LSMO film was grown on MgAl$_{2}$O$_{4}$. All films were grown via Pulsed Laser Deposition (PLD) with a KrF excimer laser at 248 nm with laser fluency of $\sim$2 Jcm$^{-2}$. The STO substrates were sourced from Shinkosa Co. LTD Japan with a manufacturers claim of less then 0.3$^{o}$ miscut, and arrived pre-etched to provide a TiO$_{2}$ terminated surface. All substrates were sonicated in isopropyl alcohol to remove organic contaminates before use. The deposition chamber base pressure was better than 5.0$\times$10$^{-7}$ Torr before the sample was heated to deposition temperature and a partial oxygen pressure was introduced. LSMO films were deposited at 700 degrees Celsius with an oxygen partial pressure of 100 mT, repetition rate of 10Hz, laser fluence of 1.8 J cm$^{-2}$ and were cooled under 300 Torr O$_{2}$ at 5 degrees per minute. BFO films were deposited at 700 degrees Celsius with an oxygen partial pressure of 5 mT, repetition rate of 20Hz, laser fluence of 1.6 Jcm$^{-2}$ and were cooled under 220 Torr O$_{2}$ at 5 degrees per minute. PZT films were deposited at 550 degrees Celcius with an oxygen partial presssure of 100 mT, repetition rate of 3 Hz, laser fluence of 1.6 Jcm$^{-2}$ and were cooled under 700 Torr O$_{2}$ at 5 degrees per minute. The growth rate of LSMO, BFO and PZT were $\sim$0.002 nm/pulse, $\sim$0.004 nm/pulse and $\sim$0.002 nm/pulse respectively. LSMO and BFO phase purity was confirmed via standard X-Ray diffraction on a Philips Xpert Pro MRD system. An example is shown in Fig. (\ref{fig:strucchar}).

The LSMO thickness was calibrated via X-Ray reflectivity measurements
on a Philips Xpert Pro MRD system. The BFO thickness was calibrated via TEM analysis, as published in\citep{doi:10.1002/adfm.201000265}. Our samples exhibit low surface roughness, less then 2.6$\textrm{\AA}$
rms, indicating smooth growth as shown in Fig.(\ref{fig:strucchar}). 

\begin{figure}[th]
\begin{centering}
\includegraphics[width=9cm]{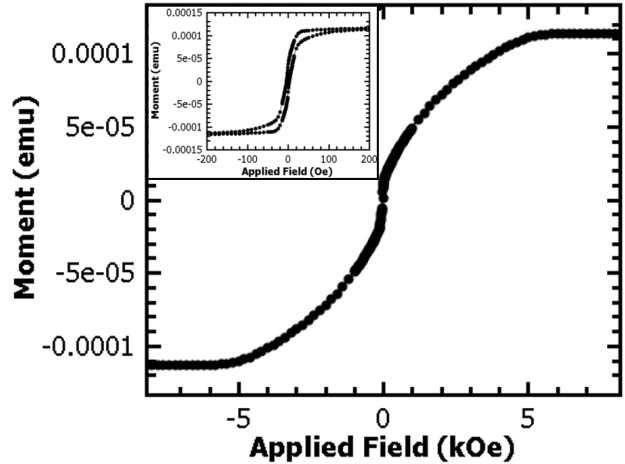}
\par\end{centering}

\caption{\label{fig:squid} SQUID data taken at room temperature with the field
applied out of the film plane, inset displays the SQUID hysteresis
at for the field applied in the plane of the film}

\end{figure}

A step pattern is seen which exists in the underlying STO substrates
and is preserved throughout the LSMO epitaxial growth. In-plane and
out-of-plane SQUID magnetometry was performed, and from this data
(seen in Fig.(\ref{fig:squid})) $\mu_{0}M_{S}-H_{oop}$ was determined
to be $\sim$0.5$\pm$0.05 T.

\section{Ferromagnetic Resonance}

\begin{figure}[th]
\centering{}\includegraphics[width=8cm]{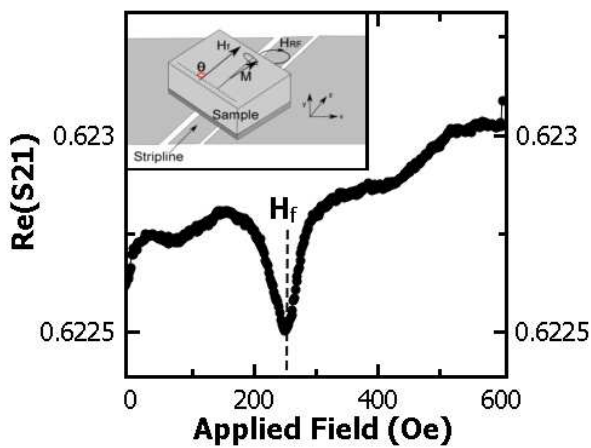}\caption{\label{fig:equipsetup}Raw data from VNA-FMR sweep for the LSMO(38.86nm)
film in the $0^{\circ}$ with a 3GHz driving microwave field. As the
applied field is swept, the Re(S21) parameter is measured (shown on
y-axis) and when resonance occurs at H$_{f}$ there is a marked change
in the Re(S21) coefficient. Inset displays a close picture of the
in-plane experimental setup with the sample sitting on top of a stripline,
M is the magnetisation precessing in response to the driving microwave
field H$_{RF}$ and the entire sample has its orientation varied by
$\theta$ with respect to the external field H$_{f}$}

\end{figure}

The FMR characterisation was done using a Vector Network Analyser
(VNA) and Field-Modulated (FM) FMR setups. The VNA-FMR is used to
obtain S21 parameters from field swept measurements as discussed in
\citep{screening-published}. It consists of a Danphysik power supply
to drive the electromagnets, and an Agilent N5230 PNA-L vector network
analyser operating in a 1-20GHz frequency range. The FM-FMR setup
uses the VNA as the microwave source, and an SRS SR850 lock-in amplifier
and HP 33120A function generator to drive the field modulated measurements.
In both cases a 0.3mm microstrip waveguide was used as the microwave
antenna source as shown in Fig.(\ref{fig:equipsetup}). The sample
is placed with the film in direct contact with the microstrip.

The in-plane FMR procedure for extracting resonance conditions is
as follows. The frequency is constant, and an external magnetic field
is swept while the S21 transmission coefficients are measured. This
procedure is repeated for several different frequencies. An example
result is shown in Fig.(\ref{fig:equipsetup}).

\section{FMR Results and discussion}

Only the FMR resonance was observed for the in-plane configuration. Lack
of FEX mode absorption may correspond to weak surface pinning
in the plane of the film. If surface pinning is weak, then the FEX
mode has a symmetric magnetisation profile across the film thickness,
producing no net dipole moment to couple to\citep{Classic_Kittel_FMR}.
In this case, only a non-uniform driving field, caused for example
by eddy currents in a conducting sample, can drive resonance\citep{kostylev-2008,screening_unpublished,PhysRev.97.1558,khivintsev:023907,PhysRev.118.658}.
Due to the low conductivity of LSMO compared to Permalloy, we do not
expect significant non-uniformity of driving field across the sample
thicknesses studied. Hence the FEX mode visibility should originate
primarily from intrinsic surface pinning. 

\begin{figure}[th]
\begin{centering}
\includegraphics[width=8cm]{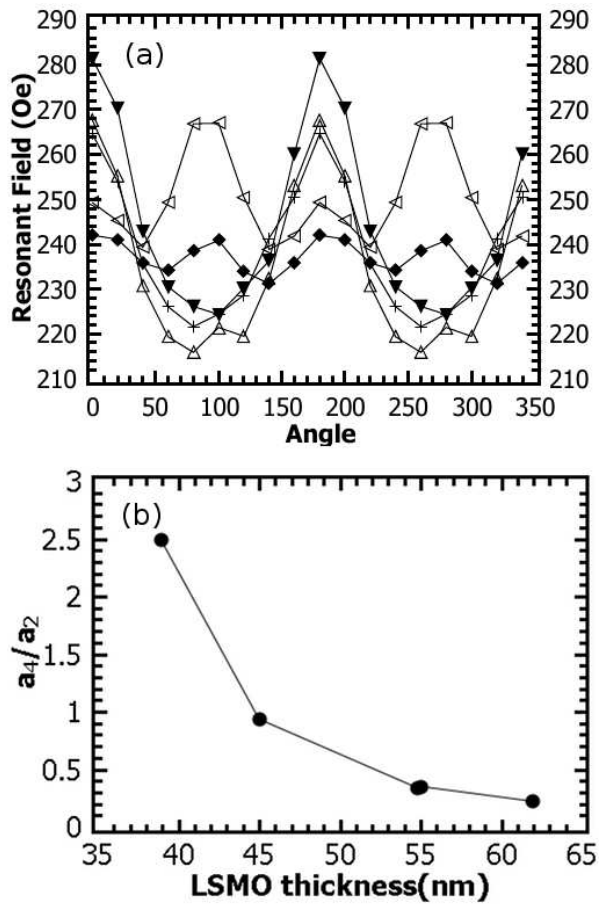}
\par\end{centering}

\caption{\label{fig:rotresonance} (a) Displayed is a plot showing the resonant
field H$_{f}$ as a function of film angle $\theta$, all experiments
used a 3GHz driving microwave field: LSMO(38.89nm) (solid diamond
line), LSMO(45nm)/PZT(20nm) (empty sideways triangle line), LSMO(54.8nm)/BFO(23nm)
(solid triangle line) , LSMO(61.9nm) (cross line), LSMO(55nm)/BFO(18nm)
(empty upright triangles). The thinnest LSMO film clearly displays
anisotropy of a biaxial character, whereas the thickest LSMO films
have a uniaxial character. Interestingly the 45nm LSMO film displays
a mixing of both anisotropy types. 
(b) shows how the ratio of
biaxial (a4) to uniaxial (a2) contributions from Eq.(\ref{eq:angular})
vary as a $\frac{1}{t}$ LSMO thickness dependence.}

\end{figure}

Magnetocrystalline anisotropies can be determined for angular measurements
in-plane, as noted above. The results for an angular study of just
the FMR mode is displayed in Fig.(\ref{fig:rotresonance}). 

The type of anisotropy seems to be dependent on the film thickness.
Biaxial and uniaxial anisotropies can be identified with the thinnest
LSMO film having a biaxial character and the thickest LSMO films having
a uniaxial character. There is also the case of the intermediate 45nm
LSMO film which displays an unequal mixing of both uniaxial and biaxial
character. Quantitative information on the form of the angular anisotropies
can be obtained by fitting to:

\begin{eqnarray}
H_{R} & = & a_{0}+a_{2}sin\left(2\theta+\phi_{1}\right)+a_{4}sin\left(4\theta+\phi_{2}\right)\label{eq:angular}\end{eqnarray}
where H$_{R}$ is the resonant applied field, $\theta$ is the film
angle with respect to the applied field, a$_{2}$ is a uniaxial anisotropy
term, a$_{4}$ is a biaxial anisotropy term. The $\phi$ are phase
shifts of the anisotropies with respect to the 0$^{o}$ measurement
direction. Examining the ratio $\frac{a_{2}}{a_{1}}$ as a function
of LSMO thickness reveals a $\frac{1}{t}$ trend as shown in Fig.(\ref{fig:rotresonance}).
This indicates that the uniaxial anisotropy dominates over the biaxial
anisotropy as the LSMO thickness increases. Furthermore, this effect
is related to the LSMO and substrate, as it does not appear to be
correlated with the capping layer.

Previous studies have noted both uniaxial and biaxial anisotropies
present in STO/LSMO films, with the biaxial anisotropy originating
from the cubic symmetries of epitaxial LSMO grown on (001) STO and
the uniaxial anisotropies originating from physical steps on the STO
surface\citep{4817880,suzuki:140,APL-1.1578711,arxiv:1005.0553v1}.
It was reported that the in-plane four-fold and two-fold anisotropies
are bulk in origin and not strongly related to interface pinning. 

It should be noted however that both anisotropies are established
during the growth process. In particular, because we measure a strong
uniaxial anisotropy for quite thick (60nm) LSMO films, it would seem
unlikely that step defects\citep{PRB.1999.58.18,PhysRevB.49.15084,NeelReducedSymmetry},
could explain these observations. The fact that this uniaxial anisotropy
is dominant in thick LSMO films suggest that some kind of bulk structure
established firstly at the step boundary, and then propagates as the
LSMO grows\citep{APL-1.1578711}.

Each of the curves in Fig.(\ref{fig:rotresonance}) has a different
dc-offset, which does not depend on LSMO thickness in a systematic
way. The most likely explanation for this is either differences in
saturation magnetisation or pinning and out of plane anisotropy originating
at the ferroelectric interface. Without the additional FEX modes for
the in-plane data it is difficult to assess the contribution made
by the ferroelectric layer to in-plane surface anisotropies.

\begin{figure}[th]
\begin{centering}
\includegraphics[width=8cm]{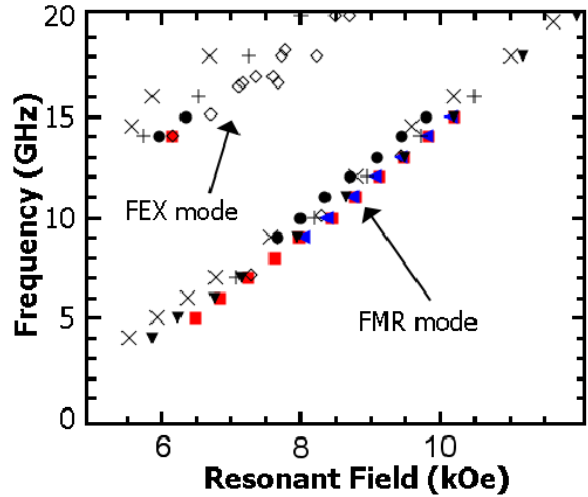}
\par\end{centering}

\caption{\label{fig:oopresonances} Out-of-plane configuration resonant field
H$_{f}$ vs driving frequency $\omega$ is shown for a variety of
different films. LSMO(30nm) (solid circles), LSMO(38.9) (solid squares
and empty diamonds), LSMO(61.9nm) (solid sideways triangles), LSMO(55nm)/BFO(17nm)
(solid down triangles), LSMO(54.8nm)/BFO(24nm) (plus symbols), LSMO(45nm)/PZT(20nm)
(cross symbols).One collection of data originates from the FMR mode
(as shown) and the other collection of points are from the FEX mode.
By fitting a linear function to these data, gyromagnetic ratio $\gamma$
and effective internal field $H_{eff}$ may be extracted.}

\end{figure}

FMR and FEX modes were seen in out-of-plane configuration measurements
for some films. The lack of an FEX mode for in-plane measurements,
and its presence in out-of-plane resonance measurements, indicate
that surface pinning is most effective in the out-of-plane direction.  When the surface pinning originates from an easy axis out-of-plane anisotropy, it has been shown that both dynamic components of magnetisation are pinned when the out-of-plane configuration, but that only one component is pinned when the magnetisation is in-plane\cite{Salansky-book}.
As shown in Fig.(\ref{fig:oopresonances}) there is a strong FMR mode
which is present in all films and a FEX mode is observed in some films.

By fitting straight lines to the data in Fig.(\ref{fig:oopresonances})
we may extract $\gamma$ and $H_{eff}$ from Eq.(\ref{eq:oopkittel-1}).
A comparison of these parameters for the different films is shown
in Tab.(\ref{tab:Heffparams}).

\begin{table*}[t]
\begin{centering}
\begin{tabular}{|c|c|c|c|c|}
\hline 
Sample & $\gamma(FMR)\times10^{10}$ & $\gamma(FEX)\times10^{10}$ & $H_{eff}(FMR)$ & $H_{eff}(FEX)$ \tabularnewline
\hline
\hline 
STO/LSMO(30nm) & 2.79 & 2.79 & -4425 & -956\tabularnewline
\hline 
STO/LSMO(38.9nm) & 2.67 & 2.67 & -4624 & -973\tabularnewline
\hline 
STO/LSMO(61.9nm) & 2.83 & - & -4891 & -\tabularnewline
\hline 
STO/LSMO(55nm)/BFO(18nm) & 2.63 & - & -4442 & -\tabularnewline
\hline 
STO/LSMO(54.8nm)/BFO(24nm) & 2.64 & 2.64 & -4423 & -437\tabularnewline
\hline 
STO/LSMO(45nm)/PZT(20nm) & 2.63 & 2.63 & -4089 & 176\tabularnewline
\hline
\end{tabular}
\par\end{centering}

\caption{\label{tab:Heffparams} Displayed are the extracted gyromagnetic ratio
$\gamma$ and effective internal field $H_{eff}$ from data shown
in Fig.(\ref{fig:oopresonances}). We attribute the distribution of
$\gamma$ to scatter in weak ferromagnetic resonance signals. Uncertainty
in $H_{eff}$ is $\pm50$ Oe, and originates from the remnant magnetic
field in our electromagnet pole pieces and the imperfect lineshapes
of the FMR signal. Also there exists large differences between the
effective fields in the single layer LSMO and capped LSMO films. In
particular, films with a more positive $H_{eff}$ must possess stronger
out-of-plane bulk anisotropies or pinning. Differences in $H_{eff}$
for the FMR and FEX modes demonstrate that interface pinning must
be playing a role in these films. For empty entries, no FEX mode was
observed.}

\end{table*}

The gyromagnetic ratio $\gamma$ is extracted from the slope of the
$\omega\left(H_{R}\right)$ data given in Fig.(\ref{fig:oopresonances}).
$H_{eff}$ is measured by the intercept with $\omega=0$ for out-of-plane
measurements. While the monolayer LSMO shows a net decrease in $H_{eff}(FMR)$
with thickness, indicating a reduction in out-of-plane anisotropies,
the addition of a ferroelectric layer significantly changes $H_{eff}$.
It should be noted that the PZT seems to much more strongly affect
the magnetic parameters then BFO.

\section{Spin Wave Stiffness}

We now discuss determination of D. We define the gap between effective
fields for the two modes $\Delta H_{eff}$ as:

\begin{eqnarray}
\Delta H_{eff} & = & H_{eff}(FEX)-H_{eff}(FMR)\label{eq:fmrgap}\\
 & = & D\left(k_{oop}^{2}(FEX)-k_{opp}^{2}(FMR)\right)\nonumber \end{eqnarray}

Eq.(\ref{eq:fmrgap}) does not contain contributions from $\mu_{0}M_{S}-H_{oop}$
, as this contributes equally to both $H_{eff}(FMR)$ and $H_{eff}(FEX)$.

Tab.(\ref{tab:H-gaps}) lists the results of this gap for films in
which the FEX mode was observed, and also for estimates of what these
values should be assuming no interface pinning and the literature
value of D\textbf{$_{lit}$= }1.7965$\times$10$^{-17}$T m$^{2}$(or
in the units for exchange stiffness D\textbf{$_{lit}=$} 104meV $\textrm{\AA}$).
We note that the spin wave constant D used in the Kittel equation
has units T m$^{2}$ and spin wave stiffness D$_{stiffness}$ possesses
units J m$^{2}$\citep{PhysRevB.76.184413}. Converting between the
two uses the following:\begin{eqnarray*}
\mathbf{D} & = & \frac{\mathbf{D_{stiffness}}}{\mu_{B}}\end{eqnarray*}

\begin{table*}
\begin{centering}
\begin{tabular}{|c|c|c|}
\hline 
 & $\bigtriangleup$H$_{eff}$ (experimental) & $\bigtriangleup$H$_{eff}$ (no pinning, D$_{lit}$)\tabularnewline
\hline
\hline 
LSMO(30nm) & 3651 & 1970\tabularnewline
\hline 
LSMO(38.9nm) & 3496 & 1171\tabularnewline
\hline 
LSMO(45nm)/PZT(20nm) & 4265 & 875\tabularnewline
\hline 
LSMO(54.8nm)/BFO(24nm) & 3986 & 590\tabularnewline
\hline
\end{tabular}
\par\end{centering}

\caption{\label{tab:H-gaps}Experimentally found $\bigtriangleup$H$_{eff}$
are shown in the left column, in units of Oe and with an error of
$\pm$50. The right hand column displays calculated $\bigtriangleup$H$_{eff}$
for the thicknesses of LSMO using the literature value of D\textbf{$_{lit}$=
}1.7965$\times$10$^{-17}$T m$^{2}$ and no interface pinning. Note
that experimental $\bigtriangleup$H$_{eff}$ pinning from different
capping layers appears significant.}

\end{table*}

\begin{table*}
\begin{centering}
\begin{tabular}{|c|c|c|c|}
\cline{2-4} 
\multicolumn{1}{c|}{} & no pinning & max single sided pinning & max double sided pinning\tabularnewline
\hline 
D$_{lit}$ & 1171 & 2343 & 3515\tabularnewline
\hline 
1.49 D$_{lit}$ & 1745 & 3491 & 5237\tabularnewline
\hline 
2.98 D$_{lit}$ & 3491 & 6983 & 10475\tabularnewline
\hline
\end{tabular}
\par\end{centering}

\caption{\label{tab:D-calculations}Theoretically calculated $\bigtriangleup$H$_{eff}$
(in Oe) for a 38.9nm LSMO film given various extreme interface pinning
conditions (top row), and different values of D (with respect to D\textbf{$_{lit}$=
}1.7965$\times$10$^{-17}$T m$^{2}$). It is important to note the
actual $\bigtriangleup$H$_{eff}$= 3496 Oe. Assuming that the entirety
of pinning originates from one interface, D\textbf{=}1.49 D$_{lit}$
as indicated in the table.}

\end{table*}

We see immediately that there is a large discrepancy between the observed
and predicted $\Delta H_{eff}$. It is not possible
from observations of two standing spin wave modes to identify separately
pinning at each interface. We thus consider two extreme situations:
complete pinning at the ferroelectric interface only (single sided
pinning) and complete pinning at both interfaces (double sided pinning)\footnote{We note that if complete pinning does occur at both interfaces, then the first exchange mode becomes symmetric and will not be excited by a uniform driving field.  In our experiment, only the conductivity would cause a non-uniform driving field, and this effect is estimated to be quite weak for LSMO.  Hence a first exchange mode which experiences double sided pinning would be extremely difficult to detect.  In our estimation of the theoretical range of D we neglect this fact.}.
D values deduced from measured $\Delta H_{eff}$ gaps in 38.9nm LSMO
films for extreme pinning conditions are given in Tab.(\ref{tab:D-calculations}).
The case where D$_{lit}$ accounts for $\Delta H_{eff}$ can only
occur for double sided pinning. In the case where extreme single sided
magnetisation pinning exists, an exchange constant value of at least
1.49$\times$ D$_{lit}$ is needed. Finally the case where $\Delta H_{eff}$
can be explained without any interface pinning is only possible for
2.98$\times$ D$_{lit}$. Most likely, a combination of both a larger
D combined with some pinning from both interfaces would account for
the D value we find. In addition, there is clearly a much greater
$\Delta H_{eff}$ gap for the films with a capping ferroelectric,
especially for the PZT capped film. This indicates a significant out-of-plane
interface pinning which originates from the ferroelectric layer.

\section{Summary}

Ferromagnetic resonance was used as a sensitive probe of both in-plane
and out-of-plane anisotropies in multilayer LSMO/BFO and LSMO/PZT
films. We have shown that some interface pinning must be playing a
role in magnetisation dynamics. Interestingly, BFO seems to have little influence
on the magnetisation of LSMO. We see no evidence of exchange bias\citep{doi:10.1021/nl801391m,adma.200502622},
though this may be because of the relatively thick LSMO layer dominating
the magnetisation dynamics. The fast growth rate of BFO in comparison
to\citep{doi:10.1021/nl801391m} may also be an important factor in
growing suitable BFO to couple to the ferromagnet. Electric fields
were applied across the BFO layer and resonance experiments were carried
out, but no shifts in FMR resonances were observed. From our data,
it is quite difficult to extract separate pinning effects at each
interface. However, a lower bound for D can be set given various assumptions
about pinning at the interface. We find that complete pinning at both
interfaces gives D for our films the same as literature D$_{lit}$. 

Uniaxial and biaxial in-plane anisotropies appear to be unrelated
to the capping ferroelectric layer are observed, and there may exist
a thickness of LSMO about which a transition between anisotropies
might take place. Our data indicates that the uniaxial contribution
to anisotropy relative to the biaxial component increases as the ferromagnet
thickness increases, for LSMO films grown on STO(100), unlike that
found for LSMO grown on other substrates\citep{4837696}.

\begin{acknowledgments}
We acknowledge the support of the ARC, University of Western Australia,
DEST and the DIISR Australia India Strategic Research Fund ST020078.
\end{acknowledgments}

%\section*{Bibliography}

%\bibliographystyle{apsrev4-1}
%\bibliography{C:/Users/Rhet/Documents/Bibliography/mybib2}

%Merlin.mbs v4.21 2009-07-09.
%

\end{document}